\documentclass[preprint]{revtex4}

\usepackage{bm}
\usepackage{graphicx}
\usepackage{epsfig}
\input{epsf}
\begin{document}
\title{Geometrically frustrated  GdInO$_3$: An exotic system to study negative thermal expansion and spin-lattice coupling}

\author{Barnita Paul, Swastika  Chatterjee and Anushree Roy}
\email{anushree@phy.iitkgp.ernet.in} \affiliation{Department of
Physics, Indian Institute of Technology Kharagpur 721302, India.}
\author{A. Midya and P. Mandal}
\email{prabhat.mandal@saha.ac.in}\affiliation{Saha Institute of
Nuclear Physics, 1/AF Bidhannagar, Calcutta 700 064, India.}
\author{Vinita Grover and A.K. Tyagi}
\affiliation{Chemistry Division, Bhabha Atomic Research Centre,
Mumbai 400085, India.}

\begin{abstract}
In this article, we report negative thermal expansion and spin
frustration in hexagonal GdInO$_{3}$. Rietveld refinement of the XRD
patterns reveal that the negative thermal expansion in the
temperature range of 50-100K stems from the triangular lattice of
Gd$^{3+}$ ions. At low temperature, the downward deviation  of the
inverse susceptibility ($\chi^{-1}$) vs. $T$ plot from the
Curie-Weiss law  indicates spin frustration which inhibits
long-range magnetic ordering down to 2K. Magnetostriction
measurements clearly demonstrate a strong spin-lattice coupling. Low
temperature anomalous phonon softening, as obtained from temperature
dependent Raman measurements, also reveals the same. Our
experimental observations are supported by first principles density
functional theory calculations of the electronic and phonon
dispersion of GdInO$_3$. The calculations suggest that the GdInO$_3$
lattice is highly frustrated at low temperature. Further, the
calculated normal mode frequencies of the Gd related $\Gamma$ point
phonons are found to depend on the magnetic structure of the
lattice, suggesting significant magneto-elastic coupling.

\end{abstract}

%
\maketitle
\def\d{{\mathrm{d}}}

\section{Introduction}

In recent years, ABO$_{3}$-type (A=rare-earth, B=transition metal)
rare-earth ferrites, manganites or nickelates find a special
interest due to their multiferroic characteristics
\cite{Huang,Giovannetti,Wang,Das,Xu,Arpita}. The role of $d$-shell
electrons of B ions, governing the multiferroic properties of these
systems, has been explored extensively. Unlike the above materials,
if B belongs to a non-transition metal ion, the electric or magnetic
properties of the system are expected to arise only from the
$4f$-shell electrons of the rare-earth ion, A. In this regard, rare
earth indates, REInO$_3$, have emerged as potential candidates for
fascinating ferroelectric memory devices
\cite{Abrahams,Tohei,Shukla,Shukla1}. The non-centrosymmetric atomic
arrangement in the hexagonal unit cell of this system gives rise to
the geometric ferroelectricity \cite{Paul}. Among all compounds in
the rare-earth indate series, GdInO$_3$ draws a special attention
because of the presence of Gd$^{3+}$ ion, which has exact
half-filled 4$f$ shell as the outermost orbital. Gd$^{3+}$ ion shows
pure spin magnetism with $\vec{L}=0$, $\vec{J}=\vec{S}=7/2$. Due to
the isotropic g-factor, specifying the magnetic moment of Gd$^{3+}$,
one expects GdInO$_3$ as a classical Heisenberg system. Furthermore,
in the literature we find that some of the Gd based compounds
exhibit the negative thermal expansion (NTE). While the mechanism of
NTE in crystalline Gd \cite{Ergin} is associated with the change in
magnetic ordering, the same is attributed to the transverse
vibrational motion of two-coordinated Pd atom in
GdPd$_{3}$B$_{0.25}$C$_{0.75}$ \cite{Pandey}. This indicates the
origin of NTE to depend on the crystalline environment of Gd ions in
a system. Along with large spontaneous polarization \cite{Paul}, the
possibility of appearance of above discussed features marks
GdInO$_{3}$ as an exotic system. Although In$^{3+}$ does not play
any direct role on spin ordering in GdInO$_{3}$, the
non-centrosymmetric distortion in the crystal structure due to large
In$^{3+}$ ion is expected to yield a complex interplay between spin
and lattice degrees of freedom in this system.

In the present work, we discuss the magnetic ordering, the crystal
structure  and the possibility of spin and lattice coupling  in
GdInO$_3$ compound. We have observed NTE in this system over the
temperature range between 50K and 100K.  The role of Gd$^{3+}$ ions
for NTE is evident from the Rietveld refinement of low temperature
X-ray diffraction (XRD) patterns. In addition, the spin frustration
in this system is confirmed from the temperature dependence of the
magnetic susceptibility below 150K. Low temperature magnetostriction
measurements indicate strong spin-phonon coupling in this system.
The anomalous softening of the phonon mode in the low temperature
range also reveals the same. Experimental observation has been
further supported by the first principles density functional theory
(DFT) calculations of the electronic and phonon dispersion in
GdInO$_3$. Our calculations find that the lattice is highly
frustrated and that there does exist a substantial amount of
spin-lattice coupling in this system. The present article is
organized as follows. In Section II, we have discussed the
experimental and computational details. Section III presents the
results regarding NTE and spin frustration in GdInO$_3$. Following
which the spin-lattice coupling in this system is discussed.
Finally, in section IV we have summarized our results.

\section{Experimental and computational details}

Bulk powder of GdInO$_{3}$ was prepared by self-assisted gel
combustion method. Stoichiometric amount of Gd$_{2}$O$_{3}$ and
In$_{2}$O$_{3}$ were dissolved in nitric acid followed by an
addition of glycine. Glycine acts as both fuel and complexing agent.
A gel was formed by evaporating the solution at the temperature of
80-100$^{\circ}$C. It was then further heated up to 250$^{\circ}$C.
The obtained powder was calcined at 550$^{\circ}$C for 1 h and
then annealed at 850$^\circ$C for 12 hrs. The
details of the synthesis procedure are reported elsewhere  \cite{Paul}.

Raman measurements on all samples were carried out using a
micro-Raman spectrometer (TRIAX550, JY, France), equipped with a
Peltier-cooled charge coupled device (CCD) as the detector. A laser
irradiation of 488 nm wavelength with 3 mW laser power was used as
an excitation source to avoid the heating of the sample. A
50L$\times$ microscope objective was used for focusing the light on
the sample. Temperature variation was carried out using a sample
stage and a temperature controller along with a liquid nitrogen pump
(THMS-600, Linkam, UK) over the temperature range 78K to 300K. From
300K to 240K, spectra were recorded at a temperature interval of 15K
whereas at 10K interval over the range of 230--180K and 5K interval
for 175K to 78K.

All magnetization measurements were carried out using SQUID VSM (Quantum Design).
We have measured the sample length change in the temperature
range 2--300 K of typical length $\sim$1 mm by the capacitive method
using a miniature tilted-plate dilatometer.
The longitudinal magnetostriction was measured
with field applied parallel to the sample length.

First principles calculations were performed within the framework of
density functional theory (DFT) \cite{Hohen,Kohn} using the
projector augmented wave (PAW)\cite{Paw,Kresse} method as
implemented in the plane-wave based VASP code
\cite{Vasp,Kresse2,Kresse3}. The exchange-correlation functional was
chosen to be the Perdew-Burke-Ernzerhof (PBE) \cite{Perdew}
implementation of the generalized gradient approximation (GGA). An
energy cut-off of 450 eV was used for plane wave expansions. To
include the strong correlation effects of \emph{4f} electrons of Gd,
we used the spin-polarized GGA plus Hubbard U (GGA+U) \cite{Hubbard}
method, as in the Dudarev's implementation \cite{Dudarev}, with U--J
=4.6 eV \cite{Topsakal}. The ionic positions as well as the lattice
parameters have been relaxed using conjugate-gradient algorithm,
until the Helmann-Feynman forces become less than 0.005 eV/{\AA}.
The energy convergence with respect to the computational parameters
was carefully examined. The $\Gamma$ point phonon frequencies have
been calculated using the density functional perturbation theory
(DFPT) \cite{Baroni} as implemented in the VASP code.

\section{Results and Discussion}

\subsection{Structural anomaly and spin-frustration in GdInO$_3$}
\begin{figure}[h!]
\centerline{\epsfxsize=4.0in\epsffile{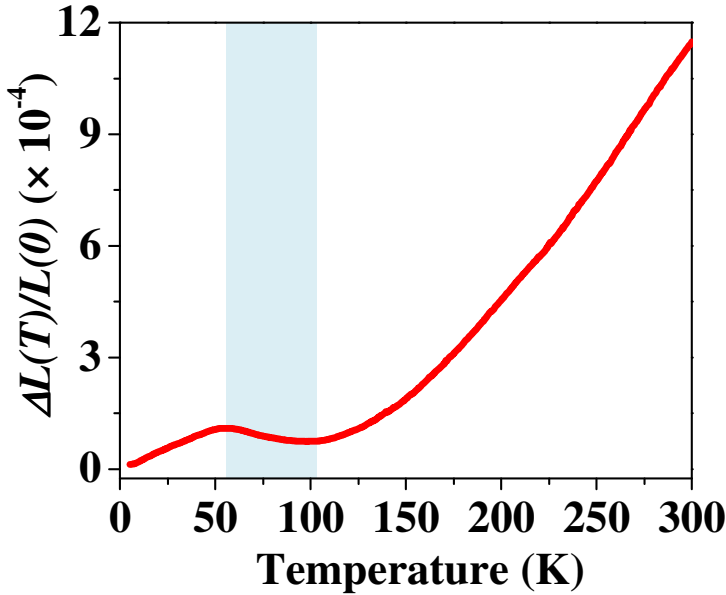}} \caption{
Temperature dependence of the relative length change $\Delta
L/L(0)$. Blue shaded zone marks the temperature range for negative
thermal expansion.} \label{expansion}
\end{figure}
Fig. \ref{expansion} displays the coupling of temperature with
lattice degrees of freedom in GdInO$_3$,  by plotting  the relative
thermal expansion of the sample length, defined by $\frac{\Delta
L}{L(0)}\equiv \frac{L(T)-L(0)}{L(0)}$, over the temperature range
between 5K and 300K.  $L(0)$ is the length of the sample at 5K (the
lowest temperature at which the measurement was carried out). Below
room temperature, the plot exhibits expected monotonic decreasing
trend till 100K. Between 100K and 50K, the value of $\Delta L$
increases with decrease in temperature, followed by a smooth
downturn upon further lowering of the temperature. The increase in
$\Delta L$ with lowering of temperature, over the range between 50 K
and 100K, indicates NTE of the system in this range of temperature.

\begin{figure}
\centerline{\epsfxsize=4.5in\epsffile{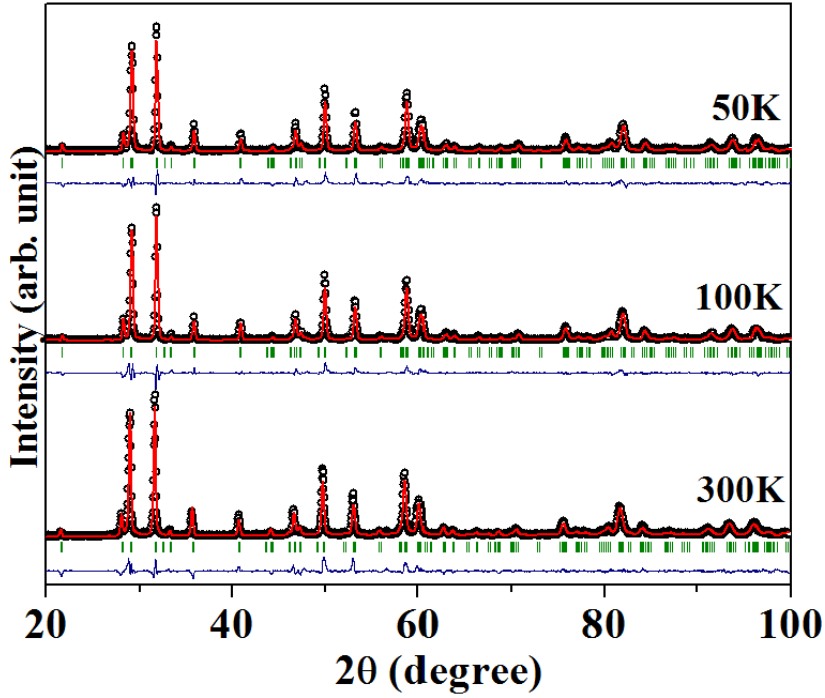}} \caption{Rietveld
refined pattern of GdInO$_3$ at 50K, 100K and 300K. The red lines
are the fitted patterns. The green bars show the positions of Bragg
reflection peaks and the blue lines are the difference between the
experimental and calculated patterns in each panel.} \label{xrd}
\end{figure}

\begin{figure}
\centerline{\epsfxsize=3.5in\epsffile{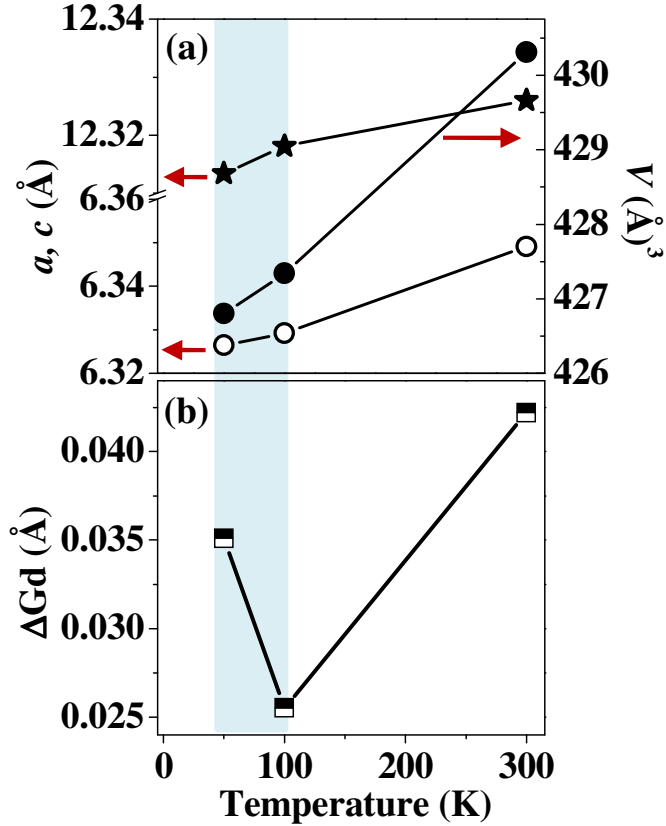}} \caption{(a)
Variation in $a (\circ)$, $c (\ast)$, and cell volume ($V, \bullet$)
and (b) $\Delta$Gd with temperature. The error bars are within the
size of the symbols.} \label{param}
\end{figure}

To decipher the origin of NTE, as seen in Fig. \ref{expansion}, we
probe the effect of temperature on the crystal structure of the
GdInO$_3$. Fig. \ref{xrd} shows the powder XRD patterns of the
compound at 50K, 100K and 300K. Rietveld refined patterns are shown
by red solid lines in the figure. All patterns could be fitted with
non-centrosymmetric P6$_3$cm space group. This rules out the
possibility of any structural phase transition to be the origin of
NTE, observed in Fig. \ref{expansion}, in the temperature range
between 50K and 100K. Fig. \ref{param}(a) plots the lattice
parameters, $a$, $c$ and unit cell volume ($V$) at three
temperatures, as obtained from the Rietveld analysis of the
diffraction data.  It is to be noted from Fig. \ref{expansion} that
for GdInO$_3$, the linear thermal expansion coefficient $\alpha_{L}$
is  of the order of $~10^{-6}$ K${^{-1}}$. Thus, it is non-trivial
to find the reflection of the expected small change in the lattice
parameters between 50K and 100K in the XRD patterns of the sample.
Therefore, instead of analyzing the change in lattice parameters, we
carefully examined bond distances between ions in the refined
structure. It is to be noted that this may reveal the contribution
of specific atomic plane in anomalous structural distortion for NTE.

\begin{figure}
\centerline{\epsfxsize6in\epsffile{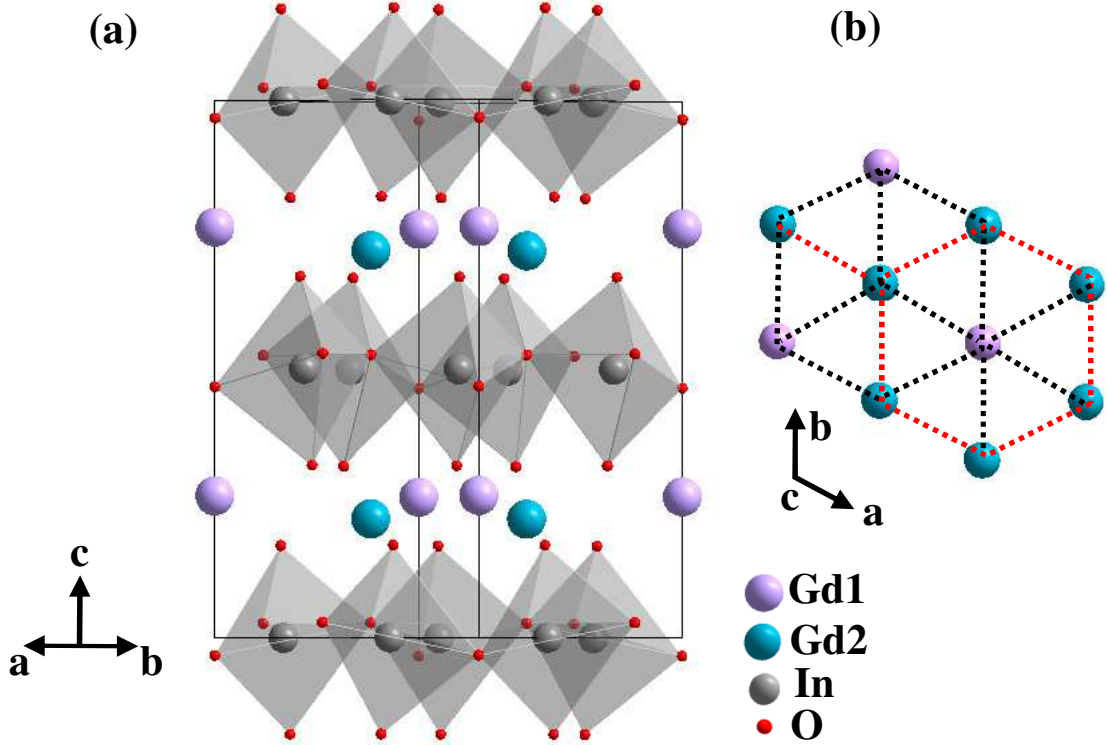}} \caption{(a) Crystal
structure of hexagonal GdInO$_3$ and (b) triangular arrangement of
Gd ions viewed along c-axis.} \label{structure}
\end{figure}

The atomic arrangement of hexagonal GdInO$_3$ unit cell with
non-centrosymmetric P6$_3$cm space group is shown in Fig.
\ref{structure}(a). Gd1 and Gd2 are two inequivalent Gd atoms with
Wyckoff positions 2a and 4b respectively. The hexagonal structure
consists of tilted InO$_5$ bipyramids with two apical (O1, O2) and
three planar oxygen ions (O3, O4, O4). Two inequivalent atomic
positions of Gd ions form a triangular lattice, as shown in Fig.
\ref{structure}(b), between two InO$_5$ bipyramidal layers.  Two
different arm lengths, $d_{1}$=Gd1-Gd2 (black dashed lines) and
$d_{2}$=Gd2-Gd2 (red dashed lines), are involved in forming the
triangular lattice. Interestingly, we find that the difference
between these two distances, ($\Delta$Gd=$d_{1}-d_{2}$), does not
change monotonically with temperature, as shown in Fig. 3(b). At 300
K the difference is 0.0422 {\AA}, which is 0.0255 {\AA} at 100 K.
The difference again increases to 0.0351 {\AA} at 50K. Comparing
Fig. 1(a) and Fig. 3(b) it appears that the above anomalous lattice
distortion in the Gd plane is reflected in NTE of GdInO$_3$ over the
temperature range between 50K and 100K.

\begin{figure}
\centerline{\epsfxsize=3in\epsffile{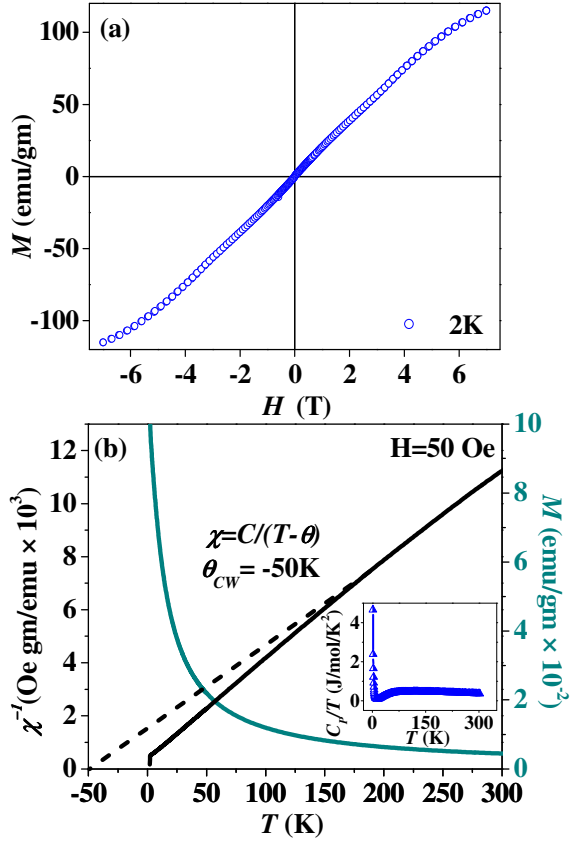}}  \caption{(a) M-H
curve at 2K, (b) Plot of $M$ and $\chi^{-1}$ vs T (solid line). The
dashed line corresponds to the Curie-Weiss law.} \label{MH}
\end{figure}

To explore the magnetic behavior, we have carried out magnetization
measurements at constant temperature($M-H$ plot at 2K) and at
constant field ($M-T$ plot at 50 Oe), Fig. \ref{MH} (a) and (b),
respectively. The absence of hysteresis loop in $M-H$ plot suggests
an absence of long range magnetic ordering in the system even down
to 2K. However, the presence of short range magnetically ordered
phase was identified from the temperature dependence of inverse
susceptibility $\chi^{-1}$ (=$H/M$) plot, shown by the solid line in
Fig. \ref{MH}(b). Though the susceptibility follows the Curie-Weiss
like behavior at high temperature, it shows a deviation from the
linearity  below 150K (the dashed line marks the expected
Curie-Weiss linear plot), indicating a strong spin fluctuation in
the paramagnetic phase of the system. The extrapolated Curie-Weiss
temperature ($\theta_{CW}$) is found to be --50K. The inset of Fig.
\ref{MH} (b) plots the variation of specific heat with temperature
($C_{p}/T$ vs. $T$). We observe a sharp upturn of the plot just
below 1.8K (Neel temperature, $T_N$). It is to be noted that
$\chi^{-1}$ shows a sharp drop just below 1.8K. Both of these
observations suggest that antiferromagnetic (AFM) ordering takes
place only below 1.8K. The ratio of $\theta_{CW}$ and $T_N$
($\theta_{CW}/T_{N}$), which estimates the degree of frustration, is
found to be sufficiently large $\sim$28.

\subsection{Spin-lattice coupling in GdInO$_3$}
\begin{figure}
\centerline{\epsfxsize=4.5in\epsffile{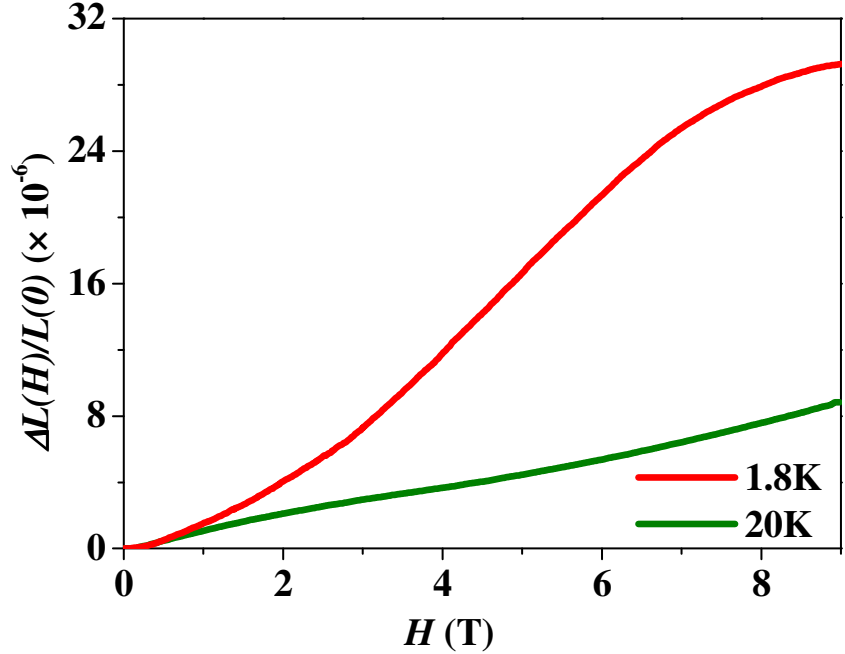}}
\caption{Magnetostriction plotted as $\Delta L(H)/L(0)$ vs $H$ at
1.8K and 20K.} \label{striction}
\end{figure}

In above discussion we confirm the presence of short range magnetic
ordering in GdInO$_3$ over a wide range of temperature. The
antiferromagnetic ordering in the system is expected only below
$T_N$=1.8K. Magnetostriction measurement is often used to study the
strength of coupling between spin and lattice. $\Delta L(H)/L(0)$
i.e. $[L(H)-L(0)]/L(0)$ plots, recorded at 1.8K and 20K are shown in
Fig. \ref{striction}. We find that at 1.8K, $\Delta L(H)/L(0)$
varies with applied magnetic field along with a change in slope at
$\sim$ 3 T. However, a very small change was observed for the same,
recorded at 20K. It is to be noted that without any coupling between
spin and lattice, we expect the $\Delta L(H)/L(0)$ plot to be
independent of temperature. Above result is a clear indication of
low temperature coupling of lattice and spin degrees of freedom in
GdInO$_3$.

\begin{figure}
\centerline{\epsfxsize=6in\epsffile{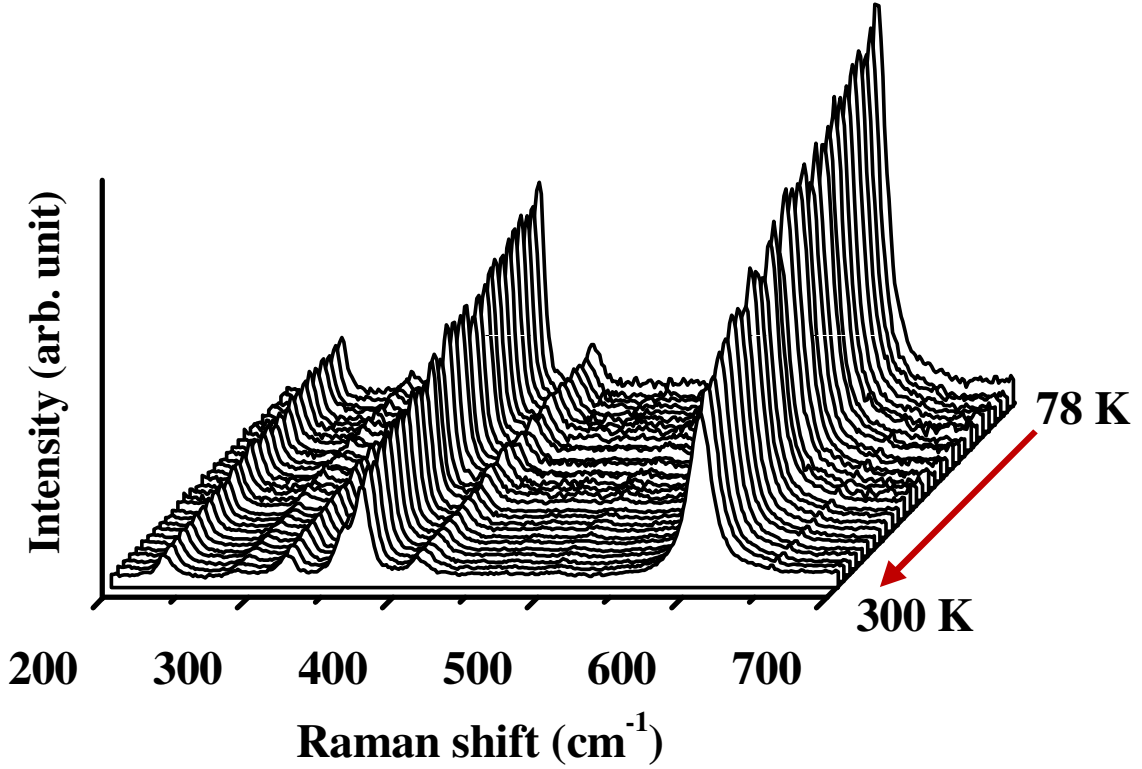}} \caption{Temperature
dependent Raman spectra of GdInO$_3$ in the spectral range of 200 to
700 cm$^{-1}$ after linear baseline correction.} \label{Raman}
\end{figure}

Furthermore, the possibility of the coupling between the spin and
lattice degrees of freedom has been probed by temperature dependent
Raman measurements over the range between 78K and 300K, and shown in
Fig. \ref{Raman}.
\begin{figure}
\centerline{\epsfxsize=6in\epsffile{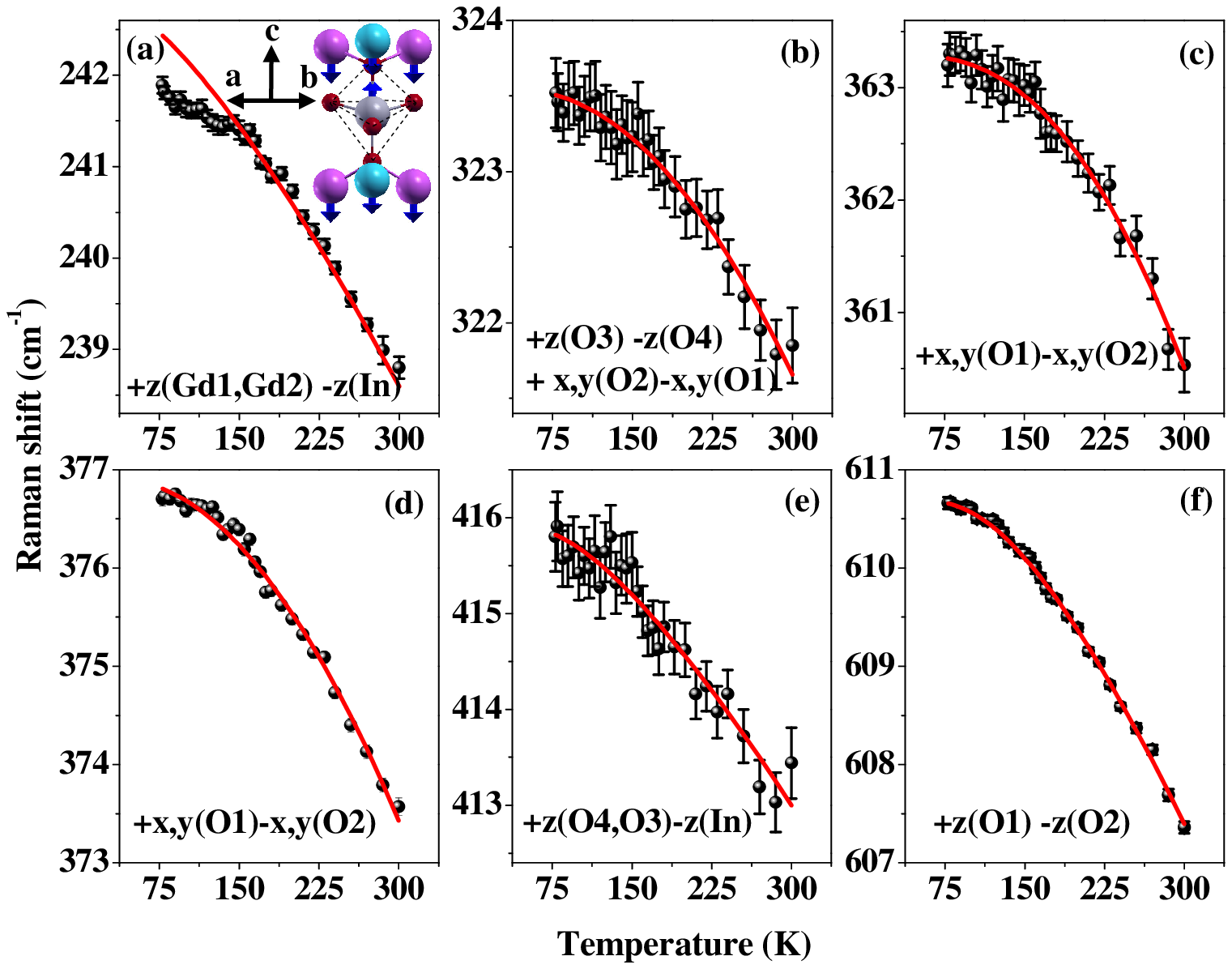}} \caption{Evolution of
Raman modes (mentioned in the text) with temperature for
GdInO$_{3}$. The inset of panel (a) schematically shows the atomic
displacement patterns of the corresponding phonon mode as mentioned
in ref \cite{Paul} involving Gd1(magenta spheres), Gd2(cyan
spheres), In(grey spheres) and O(red spheres). The blue arrows mark
the directions of the vibration of the atoms involved.}
\label{shift}
\end{figure}
All spectra were recorded over the spectral range of 200--700
cm$^{-1}$. The deconvolution of Raman spectra (after a linear
background subtraction) by seven Lorentzian profiles provides us the
evolution of Raman shift for all modes over the given range of
temperature. We consider only the prominent Raman modes around
$\sim$ 239 cm$^{-1}$, 321 cm$^{-1}$, 360 cm$^{-1}$, 373 cm$^{-1}$,
413 cm$^{-1}$ and 607 cm$^{-1}$  for further discussion. The atomic
vibration corresponding to these modes have been discussed in
details in our earlier report \cite{Paul}. Fig. \ref{shift} plots
the variation of the peak positions for the above mentioned modes
(shown by symbols) with temperature, as obtained from the
deconvolution of the spectra. The corresponding atomic vibrations of
all Raman modes from Ref. \cite{Paul} are mentioned in each panel of
Fig. \ref{shift}. The change in Raman shift with temperature at a
constant pressure can be expressed as follows
\cite{Peercy,Ravindran},
\begin{equation}
\left(\frac{d\omega}{dT}\right)_{P}=\left(\frac{\partial\omega}{\partial{T}}\right)_{V}+\left(\frac{\partial\omega}{\partial{V}}\right)_{T}\left(\frac{\partial{V}}{\partial{T}}\right)_{P}.
\end{equation}\label{omega}
Using the following definitions of Gr\"{u}neisen parameter
($\gamma$) and volume thermal expansion coefficient ($\alpha_{V}$),
\begin{equation}
\gamma=-\frac{\partial{(ln\omega})}{\partial{(lnV)}};\hspace{1mm}
\alpha_{V}=\frac{1}{V}\left(\frac{\partial{V}}{\partial{T}}\right),
\end{equation}
Eqn.1 becomes
\begin{equation}
\frac{1}{\omega}\left(\frac{d\omega}{dT}\right)_{P}=\frac{1}{\omega}\left(\frac{\partial{\omega}}{\partial{T}}\right)_{V}-\gamma\alpha_{V}.
\end{equation}
The first term in R.H.S of Eqn. 3 represents the true anharmonic
contribution of the temperature on the mode frequency due to the
anharmonic potential.  The second term represents the change in the
frequency due to the change in cell volume with temperature.  In our
case, the thermal expansion coefficient $\alpha_{V} \approx
3\alpha_{L}$ is only $\sim 10^{-6}$ (refer to Fig. \ref{expansion}),
and hence the contribution of volume contraction on the change in
the frequency of the phonon modes with lowering of temperature  can
be neglected. Thus, in Fig. \ref{shift} we analyze the
experimentally obtained data points only by considering the true
anharmonic contribution of temperature on the phonon frequency. The
solid red lines are generated by taking into account the four phonon
decay process due to the increase in anharmonicity in the
vibrational potential with temperature \cite{Balkanski}  by
following the relations
\begin{eqnarray*}
\omega_{anh}=\omega_{0}+\Delta(T),
\end{eqnarray*}
\begin{eqnarray}
\Delta(T)=A\left(1+\frac{2}{(e^{\phi/2}-1)}\right)+B\left(1+\frac{3}{({e^{\phi/3}-1})}+\frac{3}{({e^{\phi/3}-1})^2}\right),
\end{eqnarray}
where, $\omega_0$ is the phonon frequency at 0K, $\phi=
\hbar\omega{_0}/k_BT$ and $A$ and $B$ are anharmonic constants.

We find that data points for all modes, except the one at 239
cm$^{-1}$ could be fitted by a set of constants $A$, $B$ and
$\omega_0$ for each, for the whole range of temperature between 78K
and 300K (see the red solid lines in panel (b) to (f) in Fig.
\ref{shift}).  The clear change in the slope of observed phonon
frequency of the Raman mode at 239 cm$^{-1}$ near 150K compelled us
to fit the data points with free fitting parameters up to 150K using
Eqn. (1) and (2) [refer to the red solid line in the panel (a) of
Fig. \ref{shift}]. The deviation of the measured frequency of the
Raman mode near 239 cm$^{-1}$ from the expected anharmonic frequency
(solid line) indicates an additional factor to be responsible for
the anomalous shift. The atomic vibrations related to the Raman mode
at 239 cm$^{-1}$ involve vibration of Gd and In ions along the c
axis, and is shown in the inset of panel (a) in Fig. \ref{shift}. It
is to be noted that this is the only mode, under study, which
involves the vibration of Gd ions. The absence of such anomaly in
other Raman modes (panel (b) to (f) in Fig. \ref{shift}), confirms
the role of magnetic Gd plane as the origin of the anomalous phonon
softening and indicates the magneto-elastic coupling in this system.
Here we would like to mention that the plot $\Delta L(H)/L(0)$ vs.
$H$ at 20K, as observed in Fig. \ref{striction}, has much lesser
slope than that at 1.8K. From this figure it appears that the
spin-lattice coupling is much weaker at 20K. This, in a way,
contradicts the Raman results,
 which exhibit the presence of spin-phonon coupling even at 150K. This can be
justified by the fact that the magnetostriction measurement yields
an average effect over all directions. In contrary, the mode we
probe by Raman measurement involves the vibration of Gd$^{3+}$ ion
in specific direction as shown in the inset of Fig. \ref{shift}(a).

\begin{figure}
\centerline{\epsfxsize=4in\epsffile{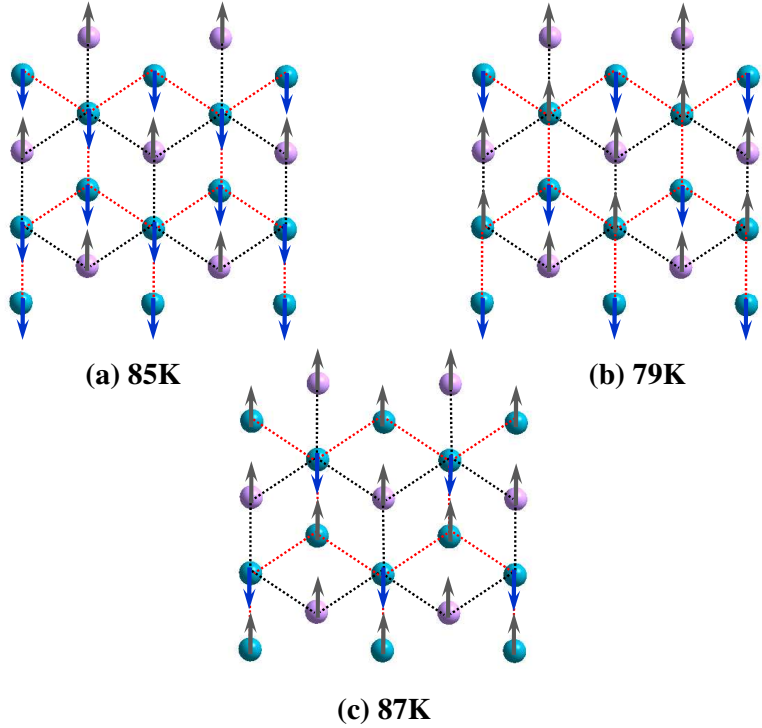}} \caption{Schematic
diagram of all AFM configurations in Gd triangular lattice. The
relative stabilization energy of  different AFM configurations with
respect to the FM configuration has been indicated at the bottom of
each AFM configuration.}\label{config}
\end{figure}

The interplay between the spin and lattice degrees of freedom in
GdInO$_3$ was further investigated  using first principles
calculations. The GdInO$_{3}$ lattice was fully optimized, starting
from the room temperature experimental data. In order to determine
the preferred magnetic orientation of Gd, we performed total energy
calculations of all possible magnetic configurations of Gd in
GdInO$_3$ lattice, as shown in Fig. \ref{config}. We find that all
antiferromagnetic configurations are energetically very close, the
difference being of the order of 0.001 eV ($\sim$10K) and that they
are higher than the ferromagnetic configuration by ~0.01 eV.
\begin{figure}
\centerline{\epsfxsize=4.75in\epsffile{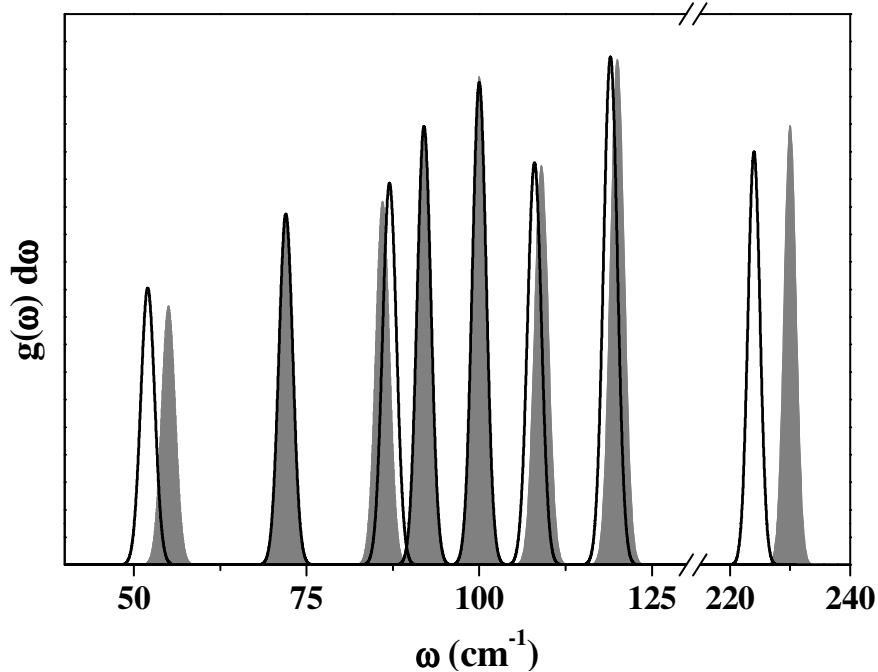}}
\caption{Calculated phonon spectra of GdInO$_{3}$ for FM (shown by
the filled area) and AFM (shown by the solid line) configuration at
0K.}\label{fm-afm}
\end{figure}
This clearly suggests that at 0K the GdInO$_{3}$ lattice is highly
frustrated. An evidence of frustration is also obtained, where we
observe a decrease in the $\Delta$Gd upon lowering of temperature.
The estimated value of $\Delta$Gd is 0.04 {\AA} at 300K and only
0.02 {\AA} at 0K. The change is very small. Nonetheless, there have
been reports on change in magnetic ordering temperature upto 30K
with only 0.1\% change in lattice parameter \cite{Singh}.

We also studied the change in the phonon frequencies  with change in
the magnetic ordering of the Gd ions in GdInO$_{3}$. For this
purpose we calculated the $\Gamma$-point phonon spectra for
ferromagnetic ordering of Gd ions and one of the many possible AFM
alignments of the Gd ions in Fig. \ref{config}. An examination of
the phonon modes indeed shows a softening as we alter the magnetic
structure of the lattice as shown in Fig. \ref{fm-afm}. The
calculated phonon mode near 230 cm$^{-1}$ (the one which
experimentally appeared at 239 cm$^{-1}$), which involves Gd ions is
found to soften
as we go from FM to AFM configuration. It is to be noted that
without spin-phonon coupling, the phonon mode frequencies would not
have been modified with a change in magnetic ordering of the system.

\section{Conclusion}

To summarize, we demonstrate NTE in hexagonal  GdInO$_{3}$ at the
temperature range of 50--100K. The corresponding change in the
crystal structure is manifested in the triangular lattice of
Gd$^{3+}$ ions. The onset of spin frustration at 150K with a large
frustration parameter hinders the long range magnetic ordering in
the system and we find an antiferromagnetic ordering only at a very
low temperature. Magnetostriction measurement and the anomalous
softening of phonon mode of Gd-related atomic vibration indicate a
spin-phonon coupling in this system. Our claims are further
supported by first principle phonon calculations.

\noindent \textbf{Acknowledgements} Authors thank Dr. P.S.Sastry,
BARC, Mumbai, for low temperature XRD measurements. AR thanks DST
and BRNS. India, for financial assistance. SC thanks DST, India for
the Inspire fellowship.


\end{document}